\documentclass[%
 reprint,
superscriptaddress,   
 amsmath,amssymb,
 aps,
prl,
]{revtex4-2}

\usepackage{graphicx}
\usepackage{dcolumn}
\usepackage{bm}

\usepackage{xcolor}

\UseRawInputEncoding
\begin{document}

\preprint{APS/123-QED}

\title{Heading-error-free optical atomic magnetometry in the Earth-field range}

\author{Rui Zhang}
\affiliation{State Key Laboratory of Advanced Optical Communication Systems and Networks, Department of Electronics, and Center for Quantum Information Technology, Peking University, Beijing 100871, China}
\affiliation{Johannes Gutenberg-Universit{\"a}t Mainz, 55128 Mainz, Germany}
 \affiliation{Helmholtz-Institut Mainz, GSI Helmholtzzentrum f{\"u}r Schwerionenforschung, 55128 Mainz, Germany}

\author{Dimitra Kanta}
\affiliation{Johannes Gutenberg-Universit{\"a}t Mainz, 55128 Mainz, Germany}
 \affiliation{Helmholtz-Institut Mainz, GSI Helmholtzzentrum f{\"u}r Schwerionenforschung, 55128 Mainz, Germany}

\author{Arne Wickenbrock}
\affiliation{Johannes Gutenberg-Universit{\"a}t Mainz, 55128 Mainz, Germany}
 \affiliation{Helmholtz-Institut Mainz, GSI Helmholtzzentrum f{\"u}r Schwerionenforschung, 55128 Mainz, Germany}

 \author{Hong Guo}
 \email{hongguo@pku.edu.cn}
\affiliation{State Key Laboratory of Advanced Optical Communication Systems and Networks, Department of Electronics, and Center for Quantum Information Technology, Peking University, Beijing 100871, China}

\author{Dmitry Budker}
\email{budker@uni-mainz.de}
\affiliation{Johannes Gutenberg-Universit{\"a}t Mainz, 55128 Mainz, Germany}
 \affiliation{Helmholtz-Institut Mainz, GSI Helmholtzzentrum f{\"u}r Schwerionenforschung, 55128 Mainz, Germany}
\affiliation{Department of Physics, University of California, Berkeley, California 94720, USA}

\date{\today}

\begin{abstract}
Alkali-metal atomic magnetometry is widely used due to its high sensitivity and cryogen-free operation. However, when operating in geomagnetic field, it suffers from heading errors originating from nonlinear Zeeman (NLZ) splittings and magnetic resonance asymmetries, which lead to difficulties in mobile-platform measurements. We demonstrate an alignment based $^{87}$Rb magnetometer, which, with only a single magnetic resonance peak and well-separated hyperfine transition frequencies, is insensitive or even immune to NLZ-related heading errors. It is shown that the magnetometer can be implemented for practical measurements in the geomagnetic environments and the photon-shot-noise-limited sensitivity reaches 9\,${\rm{fT}}/\sqrt{\rm{Hz}}$ at room temperature.
\end{abstract}

\maketitle




Highly sensitive magnetometry in the geomagnetic field is important in various applications, for example, geophysical exploration~\cite{friis2006swarm,de2007spatial,PARKER2010Geophysics,pilipenko2017ulf} and biomagnetic field detection~\cite{murzin2020ultrasensitive,Fu2020Sensitive}.
In these applications, mobile-platform-borne or even wearable magnetometry systems are especially welcome.
Scalar alkali-metal atomic magnetometers, which are based on measuring the Zeeman splitting of an alkali-metal ground state, are attractive for such tasks because of their high sensitivity and cryogen-free operation~\cite{kimball2013optical,budker2020sensing}.
However, the dependence of the (nominally scalar) magnetometer reading on the direction of the magnetic field, known as heading error, may be a limiting factor in the performance of such devices~\cite{kimball2013optical}.
Depending on the type of the atomic sensor, there are mainly three physical sources of heading error: the nonlinear Zeeman effect (NLZ) due to the coupling between electron spin and nuclear spin~\cite{Bao2018Suppression,Oelsner2019Sources,lee2021heading}, the different gyromagnetic ratios of the two ground hyperfine states due to the linear nuclear Zeeman effect (NuZ)~\cite{lee2021heading,chang2021asymmetric}, and the magnetic-field-direction-dependent light shift (LS)~\cite{Oelsner2019Sources}.
The first two effects lead to the direction-dependent asymmetry of the magnetic resonance curve, and the third one leads to the direction-dependent shift of the magnetic resonance frequency. The NLZ and LS effects are also the source of alignment-to-orientation conversion~\cite{Mozers2020Angular,Budker2000Nonlinear}.

Various methods to suppress the NLZ-related heading error have been attempted, including
synchronous optical pumping with double modulation~\cite{Seltzer2007Synchronous}, excitation of high-order atomic polarization~\cite{Acosta2008Production}, compensation with tensor light shift~\cite{Jensen2009Cancellation}, push-pull pump~\cite{Ben2010Dead}, spin-locking with synchronous optical pumping and radio-frequency
(RF)~\cite{Bao2018Suppression} or modulated optical~\cite{bao2018all} field, pumping with light of opposite circular polarization~\cite{Oelsner2019Sources}, and using a high-power pump and correcting with theoretical predictions~\cite{lee2021heading}. In some cases, heading errors due to NuZ~\cite{lee2021heading} or LS~\cite{Oelsner2019Sources} effect are suppressed together with the NLZ-related effect at the same time.

Here, we show a sensitive all-optical $^{87}$Rb magnetometer which is intrinsically free from the NLZ related heading error, with the heading error due to the NuZ and LS effects being largely suppressed.
Experiments with a room-temperature vapor cell demonstrate a photon-shot-noise-limited sensitivity at tens of \,${\rm{fT}}/\sqrt{\rm{Hz}}$ in the geomagnetic field range.
The magnetometry technique is based on the alignment magnetic resonance of the $5^2S_{1/2} F=1$ state of $^{87}$Rb confined in an antirelaxation-coated vapor cell.
As there is only a single alignment magnetic resonance in the $F=1$ ground hyperfine state, corresponding to the transition between $F=1, m=1$ and $F=1, m=-1$ states~\cite{Tailoring2011Pustelny}, magnetometry based on this resonance is free from the NLZ-related splitting and asymmetry of the magnetic resonance curve, and thus is intrinsically free from NLZ-related heading error.
In contrast to buffer-gas-filled vapor cells, there is a negligible collisional broadening of the optical transition in an antirelaxation-coated cell. As a result, the ground hyperfine states are fully resolved and the residual signal from $F=2$ is about two orders smaller than that from $F=1$ and can be neglected.
Moreover, as both the pump and probe beams used in the experiment are linearly polarized, the vector light shift of the $F=1$ alignment magnetic resonance is largely suppressed. The only light shift comes from the Zeeman-shift-related unbalanced detuning of optical transitions involving, respectively, the $F=1,\ m=1$ and $F=1,\ m=-1$ states and the imperfect polarization of the light.
An advantage of the presently introduced method for heading-error suppression for practical implementation is its particular simplicity: no special hardware or additional modulation is required.

\begin{figure*}
\centering
\includegraphics[width=2\columnwidth]{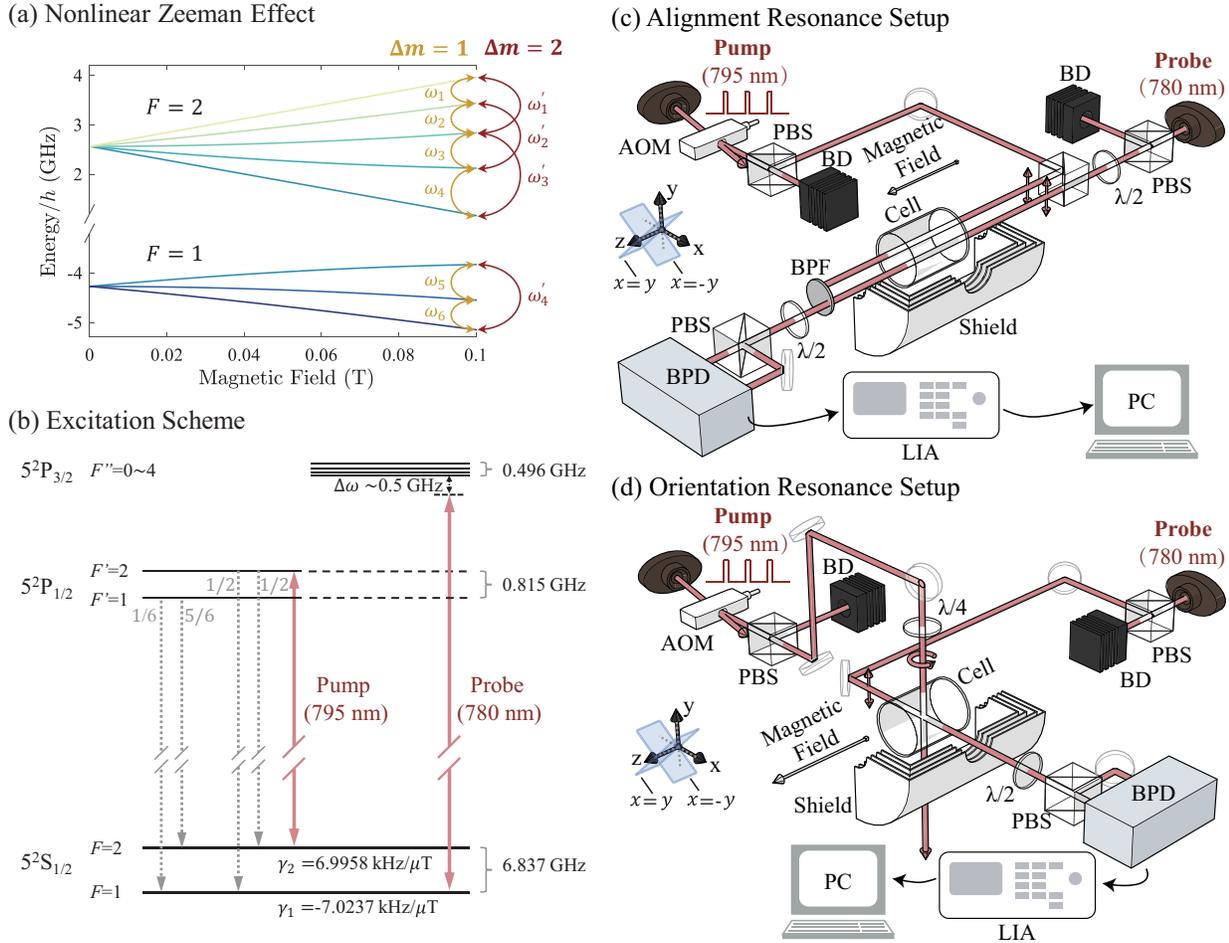}
    \caption{\textbf{
    Schematic of NLZ-heading-error-free magnetometer. }
    \textbf{(a)}
  Nonlinear Zeeman effect of $^{87}$Rb ground state $5^2{\rm{S}}_{1/2}$ and corresponding magnetic resonance frequencies. In the $F=2$ ($F=1$) hyperfine state, there are four (two) different $\Delta m=1$ magnetic resonance frequencies, labelled with $\omega_1$ to $\omega_4$ ($\omega_5$ to $\omega_6$), respectively, and 3 (1) different $\Delta m=2$ magnetic resonance frequencies, labelled $\omega_1'$ to $\omega_3'$ ($\omega_4'$), respectively.
    \textbf{(b)} Energy levels of $^{87}$Rb atoms and excitation scheme. Pump beam is a 795 nm laser beam exciting the $5^2{\rm{S}}_{1/2} F=2$ to $5^2{\rm{P}}_{1/2} F'=2$ transition, which generates atomic spin polarization in the $5^2{\rm{S}}_{1/2} F=1$ state via repopulation pumping; the probe beam is a 780\,nm laser tuned to the low-frequency side relative to the $5^2{\rm{S}}_{1/2} F=1$ to $5^2{\rm{P}}_{3/2} F''=0$ transition.
    \textbf{(c)}
    Alignment Resonance Setup.
    Pump, a linearly polarized 795 nm laser beam used to generate the atomic alignment polarization;
    Probe, a linearly polarized 780 nm laser beam used to detect the Larmor precession of the atomic alignment polarization via optical rotation;
    AOM, acousto-optic modulator used to pulse the pump beam;
    PBS, polarizing beam splitter;
    BS, beam splitter;
    $\lambda/2$, half-wave plate; $\lambda/4$, quarter-wave plate;
    BPD, balanced photodiode;
    BPF, bandpass filter with central wavelength of 780 nm, which is used to prevent the pump beam from entering the BPD;
    BD: beam damp;
    LIA, lock-in amplifier; PC, personal computer.
    \textbf{(d)}
    Orientation Resonance Setup.
    Pump, a circularly polarized 795 nm laser beam used to generate the atomic orientation polarization;
    Probe, a linearly polarized 780 nm laser beam used to detect the Larmor precession of the atomic orientation polarization via optical rotation; Other labels are the same as those in (c).
 }\label{Fig:Level_Schematic_Apparatus}
\end{figure*}

\begin{figure}
\centering
\includegraphics[width=1\columnwidth]{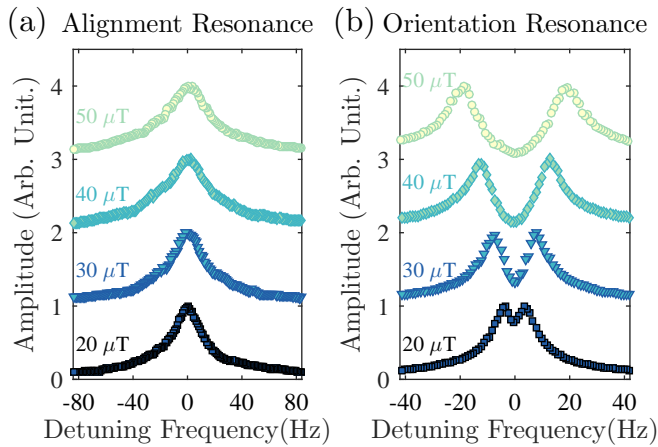}
    \caption{
    \textbf{Magnetic resonances of alignment and orientation polarization with background magnetic field of different strengths.}
    \textbf{(a)} Alignment resonance.
    \textbf{(b)} Orientation resonance.
    Background magnetic field is set along $z$ direction, with strength ranging from 20\,$\mu{\rm{T}}$ to 50\,$\mu{\rm{T}}$.
    Different magnetic resonances are shifted vertically for clarity.
    Since the alignment-resonance frequency is about twice that of orientation, the scale of detuning in (a) is also twice that of (b).
 }\label{Fig_2}
\end{figure}

\begin{figure}
\centering
\includegraphics[width=1\columnwidth]{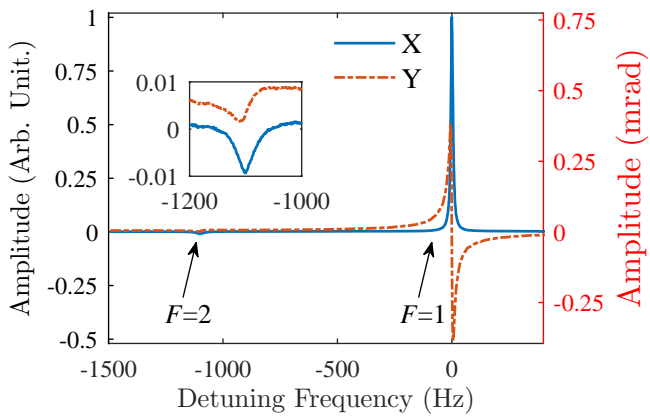}
    \caption{
    \textbf{Typical $F=1$ and $F=2$ alignment magnetic resonance.}
    Background magnetic field is set to 20\,$\mu{\rm{T}}$ along the $z$ direction. The insertion shows details about the $F=2$ resonance. This magnetic resonance is averaged from four scans.
 }\label{Fig_3}
\end{figure}

\begin{figure*}
\centering
\includegraphics[width=2\columnwidth]{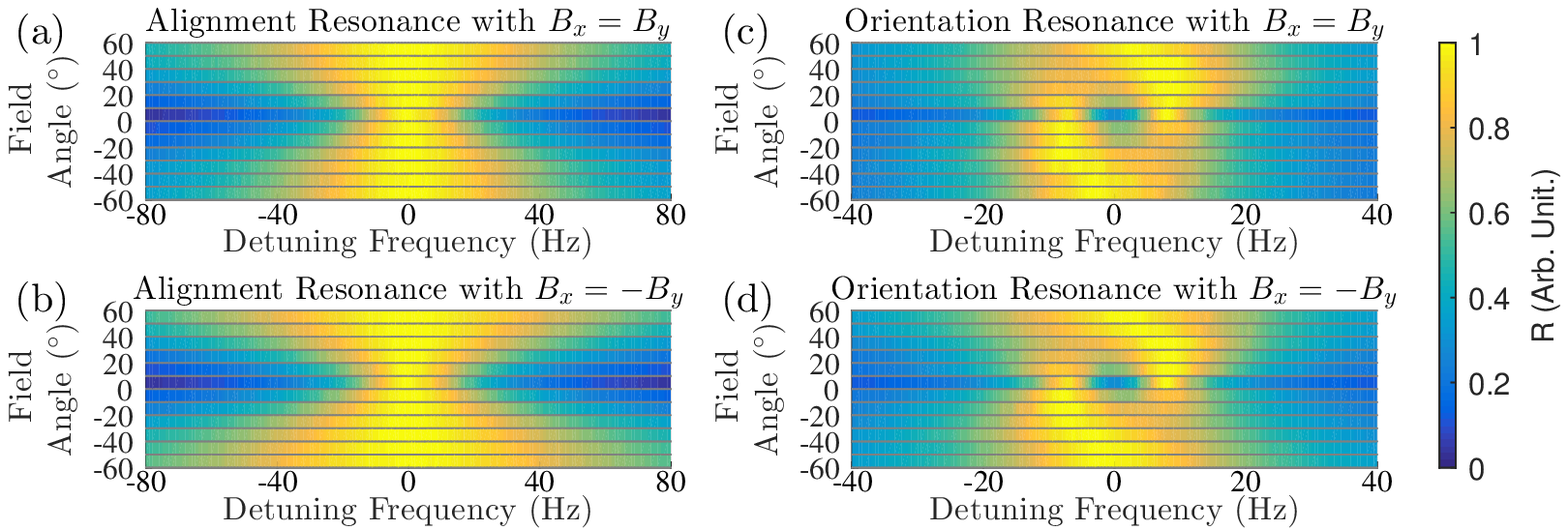}
    \caption{
    \textbf{Magnetic resonances of alignment and orientation polarization with background magnetic field along different directions.}
    \textbf{(a)} Alignment resonance with $B_x=B_y$.
    \textbf{(b)} Alignment resonance with $B_x=-B_y$.
    \textbf{(c)} Orientation resonance with $B_x=B_y$.
    \textbf{(d)} Orientation resonance with $B_x=-B_y$.
The strength of the background magnetic field was kept at 30\,$\mu{\rm{T}}$. The angle between the background magnetic field and the $z$ direction is represented as the vertical axis. The broadening of the magnetic resonances at a larger field angle is due to the increased magnetic field gradients. Since this broadening does not affect our conclusion about resonance symmetry, and that the compensation of such gradients along different directions is complex, these gradients are not compensated during this experiment.
}\label{Fig_4}
\end{figure*}

The NLZ effect of an alkali ground state, such as the $5^2{\rm{S}}_{1/2}$ state for $^{87}$Rb shown in Fig.\,\ref{Fig:Level_Schematic_Apparatus}, is described by the Breit-Rabi formula~\cite{steck2010rubidium}.
The energies of Zeeman sublevels have a nonlinear dependence with respect to the magnetic field strength, and thus, the intervals between adjacent Zeeman sublevels become unequal as the field increases. As a result, in the $F=2$ ($F=1$) hyperfine state, there are four (two) different $\Delta m=1$ magnetic resonance frequencies, which leads to asymmetric broadening or splitting of the magnetic resonance curve when the background field is in the geomagnetic field range~\cite{Bao2018Suppression,Oelsner2019Sources,lee2021heading}.
As the populations and transitions matrix elements for different Zeeman sublevels change  with the magnetic field direction,
the amplitudes of  different components of the magnetic resonance change as well, which leads to the magnetic-field-direction dependent asymmetry of the overall magnetic resonance and gives rise to heading errors~\cite{Bao2018Suppression,Oelsner2019Sources,lee2021heading}.
In contrast to the $\Delta m=1$ magnetic resonance, there is only a single $\Delta m=2$ resonance in the $F=1$ hyperfine state. Such a single magnetic resonance is intrinsically free from the NLZ related splitting and asymmetry of the resonance curve, and thus is free from the NLZ related heading error.

In order to generate and measure atomic polarization in the $F=1$ hyperfine state, see Fig.\,\ref{Fig:Level_Schematic_Apparatus}(b), we pump with 795\,nm laser light resonant with the $^{87}$Rb D1 transition ($5^2{\rm{S}}_{1/2} F=2\rightarrow5^2{\rm{P}}_{1/2} F'=2$) and probe with 780\,nm laser light tuned to the low-frequency side of the $^{87}$Rb D2 transition (about 0.5\,GHz from the $5^2{\rm{S}}_{1/2} F=1\rightarrow 5^2{\rm{P}}_{3/2} F''=0$ transition).
The pump beam generates atomic polarization in the $F=1$ state via repopulation pumping~\cite{Happer1972Optical}, and the polarization is monitored by detecting the optical rotation of the probe light.
To achieve magnetic resonance, the pump beam is modulated at two times the Larmor frequency of the $F=1$ hyperfine state. In the geomagnetic field range, the splitting between the $F=1$ and $F=2$ resonance frequencies is in the kHz range, see Fig.\,\ref{Fig_3}.
As this splitting is much larger than the relaxation rate of the ground-state polarization, the $F=2$ state will not be trapped in the dark state and the pumping process is efficient.
Since the pump and probe beam are resonant with transitions starting from different ground hyperfine states, such a technique constitutes indirect pumping which does not cause power broadening of the magnetic resonance~\cite{Chalupczak2012Room,Gartman2015Amplitude}.

The experimental arrangement for measuring the alignment magnetic resonance is shown in Fig.\,\ref{Fig:Level_Schematic_Apparatus}(c).
Both the pump and the probe beam are linearly polarized and propagate through the atomic vapor cell along the $z$ direction. The pump beam is modulated with an acousto-optic modulator (AOM) at around two times the Larmor frequency. The corresponding alignment magnetic resonances with background magnetic field set along the $z$ axis are shown in Fig.\,\ref{Fig_2}(a).
We also built an orientation-magnetic-resonance setup to compare the alignment/orientation resonances, as shown in Fig.\,\ref{Fig:Level_Schematic_Apparatus}(d).
The pump (probe) beam is circularly (linearly) polarized and propagates through the atomic vapor cell along the $y$ ($x$) direction.
In this case, the pump beam is modulated at around the Larmor frequency.
The orientation magnetic resonances with background magnetic field set along the $z$ axis are shown in Fig.\,\ref{Fig_2}(b).
In order to maintain the consistency of experimental conditions, the two setups share the same vapor cell, and the pump and probe beams of these two setups are derived from the same pump and probe lasers, respectively.

 A comparison of the magnetic field dependences of alignment and orientation magnetic resonances is shown in Fig.\,\ref{Fig_2}, in which the background magnetic field is set along the $z$ direction, with strength varying from 20\,$\mu{\rm{T}}$ to 50\,$\mu{\rm{T}}$ (data are shifted vertically for clarity).
The optical power of the pump and probe beam in the alignment (orientation) experiment is 100\,$\mu{\rm{W}}$ (50\,$\mu{\rm{W}}$) and 300\,$\mu{\rm{W}}$ (15\,$\mu{\rm{W}}$), respectively. These parameters are chosen to produce relatively strong signals with minimal power broadening of the magnetic resonances. A $\partial B_z/\partial z$ gradient coil is used to compensate the magnetic field gradient. We find that compensating this gradient is sufficient for the purpose of this work.
The alignment signal is always a single Lorentzian peak, with the central frequency at $2\omega_{\rm{L}}$, corresponding to the $\Delta m=2$ magnetic resonance between $F=1, m_{\rm{F}}=-1$ and $F=1, m_{\rm{F}}=1$ state, while the orientation signal consists of two peaks with increased splitting as the magnetic field is increased.
Both the alignment and orientation resonances slightly broaden at stronger magnetic fields due to residual magnetic gradients.

Another source of heading error comes from the different Larmor frequencies of the $F=1$ and $F=2$ ground hyperfine states, the difference of which is around 27.9~Hz/$\mu$T. In the geomagnetic field range of tens of $\mu$T, the difference is in the kHz range.
Taking a magnetic resonance at $20\,\mu$T for example, as shown in Fig.\,\ref{Fig_3}, the amplitude of the $F=2$ resonance is about 100 times smaller than that of the $F=1$ resonance. Considering the large frequency difference between these two resonances, the influence from the $F=2$ resonance on the $F=1$ Larmor frequency only leads to heading error on the order of tens of fT (see supplementary material for more details). 
The light shift due to the probe beam is also a source of heading error. When the background magnetic field is not aligned with the polarization of the probe beam, this beam generally contains the $\pi$ components together with the $\sigma^+$ and $\sigma^-$ components with equal strength. As their light shifts on the $F=1,\ m=-1$ and $F=1,\ m=1$ states are almost identical, that of the alignment resonance is largely suppressed~\cite{le2013dynamical}. However, there is still residual light shift due to the Zeeman-effect-related unbalanced detuning of optical transitions involving, respectively, the $F=1,\ m=1$ and $F=1,\ m=-1$ states, which is on the order of several pT, according to simulation based on the ADM package~\cite{ADM2022}. If the optical polarization of the probe beam is actively rotated to make it always perpendicular to the magnetic field, then the light shift will only lead to a constant bias, rather than a heading error. This method will also help to build a dead-zone-free magnetometer~\cite{Wu2015deadzone}.
In principle, there may also exist systematic effects due to interference effects including those mediated by radiative polarization transfer~\cite{Horbatsch2010Shifts,Marsman2017Interference}. The presence of such effects can be identified by measuring the heading error as a function of the light power (particularly, the pump power) and completely eliminated using a ``free-decay'' protocol, where atomic evolution occurs in the absence of applied light between pump and probe light pulses.

The magnetic-field-direction dependences of alignment and orientation resonances are shown in Fig.\,\ref{Fig_4}.
 The magnetic field strength is kept at 30\,${\mu}\rm{T}$ and the field direction is rotated in the $B_x=B_y$ and the $B_x=-B_y$ planes, respectively (to intuitively illustrate how the magnetic field is rotated, the $x=y$ and $x=-y$ planes are shown in Fig.\,\ref{Fig_4}(c) and (d), nearing the coordinate system). The relative heights of the two Lorentz peaks in the orientation magnetic resonance are dependent on the magnetic field direction, which leads to an asymmetry and a central-frequency shift of the magnetic resonance when the field is not along the $z$ direction. This gives rise to the NLZ related heading error.
 In contrast, the alignment resonance is always symmetric. This means that magnetometry based on this alignment resonance is free from the NLZ related heading error. The broadening of both the orientation and the alignment magnetic resonance at a larger field angle is due to the increased magnetic gradient, as the $x$/$y$-directed magnetic field produced by the magnetic coil is less uniform than the z-directed magnetic field.

In the geomagnetic field range, the estimated photon-shot-noise-limited sensitivities~\cite{Budker2000Sensitive,Acosta2006Nonlinear,Zhivun2016Vector} of the alignment-based magnetometry are in the tens of ${\rm{fT}}/\sqrt{{\rm{Hz}}}$ range (see supplementary material for more details).
The best sensitivity is about 9\,${\rm{fT}}/\sqrt{{\rm{Hz}}}$.
 When the background field gets larger, there is a degradation of the sensitivity. One possible reason for this is the increased magnetic field gradient.

To conclude, we demonstrate a sensitive heading-error-free scalar magnetometer which can work in geomagnetic environment.
This magnetometer is based on the ${\Delta}m=2$ magnetic resonance in the $^{87}$Rb $F=1$ ground hyperfine state.
In contrast to conventional alkali-metal magnetometry, where the magnetic resonance curve is split and distorted in the geomagnetic field, the magnetic resonance demonstrated here is only a single Lorentzian peak and so is free from NLZ induced splitting and asymmetry.
For our magnetometer, the photon-shot-noise limited sensitivity can reach $9\,{\rm{fT}}/\sqrt{{\rm{Hz}}}$, with the vapor cell working at room temperature.
The sensitivity could be further improved by heating the cell to increase the atomic vapor density~\cite{Li2017Characterization}.
This scheme is also effective at suppressing the heading error due to the NuZ and LS effects. Due to the fully resolved ground hyperfine states in the antirelaxation-coated cell,  the  residual  signal  from $F=2$ is relatively small and thus its influence is at most at the tens of fT level, which is at the limit of the sensitivity of this magnetometer for 1~Hz bandwidth. 
Moreover, as there is only linearly polarized light used in this magnetometer, the vector light shift is also largely suppressed~\cite{le2013dynamical}, which is another possible source of heading error.
Considering that the remaining tensor light shift will not change the frequency difference between $m=\pm 1$ magnetic sublevels in the ground $F=1$ system, i.e., the central frequency of the desired $\Delta m=2$ magnetic resonance, this scheme is promising for more accurate magnetometry.
It should be noted that the magnetic resonance frequency itself still has cubic correction in the strength of the magnetic field. Finally, a similar method can be implemented with other alkali metals that have a $F=1$ ground level, such as $^{39}$K, $^{41}$K and $^{23}$Na.

\section*{Acknowledgement}
The authors thank Brian~B.~Patton, Simon~M.~Rochester, Sheng~Li, Oleg~Tetriak and D.~C.~Hovde for helpful discussions.
This work was supported in part by the DFG Project ID 390831469: EXC 2118 (PRISMA+ Cluster of Excellence), the European Research Council (ERC) under the European Union Horizon 2020 Research and Innovation Program (grant agreement No. 695405), the DFG Reinhart Koselleck Project and the German Federal Ministry of Education and Research (BMBF) within the Quantumtechnologien program (FKZ 13N15064).
RZ acknowledges support from the China Scholarship Council (CSC) enabling his research at the Helmholtz-Institut Mainz  and thanks Wei Xiao for proofreading the manuscript.
DK acknowledges Martin Engler's help with the LabVIEW code.

\section*{Supplementary Material}

\subsection{Estimation of NuZ-related Heading Error}
As shown in Fig.\,3 in the Main Text, $F=1$ resonance frequency at 20\,$\mu{\rm{T}}$ can be determined from either the peak frequency of the demodulated X signal or the zero-crossing frequency of the demodulated Y signal.
The peak frequency of X (zero-crossing frequency of Y), however, is also influenced by the $F=2$ resonance, because the X (Y) contribution from the $F=2$ leads to a sloping background (residual background) at around the $F=1$ resonance frequency.
Since the $F=2$ resonance amplitude is influenced by the magnetic field direction, this background is also magnetic-field-direction related and thus is a source of heading error.

Assuming the contributions of $F=\alpha$ resonance ($\alpha$=1 or 2) to the demodulated X and Y data have the Lorentzian form of
\begin{equation}
\label{Eq_Demodulated_X_Y}
\begin{split}
X_\alpha&={\rm{Re}}\Bigl(\frac{A_\alpha\delta\nu_\alpha}{i(\nu_\alpha-\nu_{\rm{m}}) +\delta\nu_\alpha}\Bigr)=\frac{A_\alpha\delta\nu_\alpha^2}{(\nu_\alpha-\nu_{\rm{m}})^2 +\delta\nu_\alpha^2},\\
Y_\alpha&={\rm{Im}}\Bigl(\frac{A_\alpha\delta\nu_\alpha}{i(\nu_\alpha-\nu_{\rm{m}}) +\delta\nu_\alpha}\Bigr)=\frac{A_\alpha\delta\nu_\alpha(\nu_\alpha-\nu_{\rm{m}})}{(\nu_\alpha-\nu_{\rm{m}})^2 +\delta\nu_\alpha^2},\\
\end{split}
\end{equation}
in which $A_{\alpha}$, $\nu_{\alpha}$ and $\delta \nu_{\alpha}$ are the amplitude, central frequency and the half width at half maximum (HWHM) linewidth of the $F=\alpha$ resonance, while $\nu_{\rm{m}}$ is the modulation frequency of the pump beam. (Note that the lineshape for the $Y_{2}$ case is not quite dispersive in practice, see Fig.\,\ref{Fig_3} in the Main Text; however, the assumed signal dependence scaling as inverse detuning is safe for estimating the upper limits on the parasitic shift.)
The presence of the $X_2$ and $Y_2$ term leads to a shift of the effective central frequency of the magnetic resonance around $\nu_1$.

Taking the $X$ signal as an example, the effective peak frequency is the modulation frequency $\nu_{\rm{m}}$ at around $\nu_1$ which leads to a zero slope of the overall $X$ signal, e.g.,
\begin{equation}
\label{Eq_Demodulated_X_Shift_1}
\begin{split}
\Bigl(\frac{\partial X_1}{\partial \nu_{\rm{m}}}+\frac{\partial X_2}{\partial \nu_{\rm{m}}} \Bigr)\Bigr|_{\nu_{\rm{m}}\sim\nu_1}=0.
\end{split}
\end{equation}
According to Eq.~(\ref{Eq_Demodulated_X_Y}), the slopes of $X_1$ and $X_2$ data in the vicinity of the resonance, i.e., $\nu_1=\nu_{\rm{m}}$, have the form of
\begin{equation}
\label{Eq_Demodulated_X_slope}
\begin{split}
\frac{\partial X_1}{\partial \nu_{\rm{m}}}\Bigr|_{\nu_{\rm{m}}\sim\nu_1}
&\approx\frac{2A_1(\nu_1-\nu_{\rm{m}})}{\delta\nu_1^2},\\
\frac{\partial X_2}{\partial \nu_{\rm{m}}}\Bigr|_{\nu_{\rm{m}}\sim\nu_1}
&\approx\frac{2A_2\delta\nu_2^2}{(\nu_2-\nu_{1})^3},\\
\end{split}
\end{equation}
so the shift of the peak frequency of $X$ is
\begin{equation}
\label{Eq_Demodulated_X_Shift_2}
\begin{split}
\Delta\nu_{1X}\approx \frac{A_2\delta\nu_1^2\delta\nu_2^2}{A_1(\nu_2-\nu_{1})^3}.
\end{split}
\end{equation}

As for the $Y$ signal, the effective zero-crossing frequency is the modulation frequency $\nu_{\rm{m}}$ at around $\nu_1$ which leads to a zero value of the overall $Y$ signal, e.g.,
\begin{equation}
\label{Eq_Demodulated_Y_Shift_1}
\begin{split}
(Y_1 + Y_2)     \bigr|_{\nu_{\rm{m}}\sim\nu_1}=0.
\end{split}
\end{equation}
According to Eq.~(\ref{Eq_Demodulated_X_Y}), the $Y_1$ and $Y_2$ data in the vicinity of the resonance, i.e., $\nu_1=\nu_{\rm{m}}$, have the form of
\begin{equation}
\label{Eq_Demodulated_Y_slope}
\begin{split}
Y_1\Bigr|_{\nu_{\rm{m}}\sim\nu_1}
&\approx\frac{A_1(\nu_1-\nu_{\rm{m}})}{\delta\nu_1},\\
Y_2\Bigr|_{\nu_{\rm{m}}\sim\nu_1}
&\approx\frac{A_2\delta\nu_2}{(\nu_1-\nu_2)},\\
\end{split}
\end{equation}
so the shift of the zero-crossing frequency of $Y$ is
\begin{equation}
\label{Eq_Demodulated_Y_Shift_2}
\begin{split}
\Delta\nu_{1Y}\approx \frac{A_2\delta\nu_1\delta\nu_2}{A_1(\nu_2-\nu_{1})}.
\end{split}
\end{equation}

%
%
%
%
%
%

As shown in Fig.\,3, the amplitude of the $F=2$ resonance is only about 1\% that of the $F=1$ resonance, and $\delta\nu_1$,  $\delta\nu_2$ and $|\nu_1-\nu_2|$ are 5.5\,Hz, 19.2\,Hz and 1100\,Hz, respectively.
As a result, $\Delta\nu_{1X}$ and $\Delta\nu_{1Y}$ are about 84\,nHz and 0.96\,mHz, respectively.
 As the gyromagnetic ratio is $\gamma_1\approx -7$\,Hz/nT and the alignment-resonance frequency is two times the Larmor frequency, these 84\,nHz and 0.96\,mHz frequency shifts correspond to measurement errors of 6\,aT and 69\,fT, respectively.
The heading error due to the $X_2$ or $Y_2$ contributions is on the same order of magnitude with this measurement error, so the NuZ-related heading error is dependent on whether $X$ or $Y$ signal is used for measurement, and will not exceed tens of fT level even if the $Y$ data are used to determine the measurement result.



\subsection{Sensitivity Estimation}

The sensitivity of an optically pumped magnetometer is limited by two fundamental quantum noises, which are the spin-projection noise in measuring the pointing angle of the atomic spins and the photon shot noise in measuring the polarization angle of the probe beam, respectively~\cite{kimball2013optical}.
The spin-projection-noise-limited sensitivity measured in noise spectral density is given by~\cite{budker2020sensing,Zhivun2016Vector}
\begin{equation}
\label{Eq_Sensitivity_Estimation_0}
\delta B \approx \frac{1}{\gamma_1 \sqrt{2 F}}\frac{1}{\sqrt{N\tau_2}},
\end{equation}
where $\gamma_1\approx-7\,{\rm{Hz/nT}}$ is the gyromagnetic ratio of the $F=1$ ground state, $F=1$ is the total angular momentum of the system, $N$ is the atoms in the vapor cell and $\tau_2$ is the lifetime of the atomic polarization.
Given that the $N\approx9\times 10^{10}$ in the room temperature vapor cell with a radius of 2 cm and length of 5 cm and $\tau_2$ is about tens of ms, the spin-projection-noise-limited sensitivity is in the ${\rm{fT}}/\sqrt{\rm{Hz}}$ range, which is one or two orders of magnitude smaller than the photon-shot-noise limited sensitivity of our current setup. As a result, we focus on the photon-shot-noise limited sensitivity, shown as below.

The photon-shot-noise limited sensitivity of an optical-rotation magnetometer is
given by~\cite{Budker2000Sensitive,Acosta2006Nonlinear}
\begin{equation}
\label{Eq_Sensitivity_Estimation_1}
\delta B \approx \Bigl(\frac{\partial \phi}{\partial B}\Bigl|_{\nu_{\rm{m}}=\kappa\gamma_1 B}\Bigr)^{-1} \delta \phi,
\end{equation}
where $\phi$ is the optical rotation amplitude of the probe beam and is a function of $B$ (magnetic field) and $\nu_{\rm{m}}$ (modulation frequency of the pump beam), $\kappa$ is the rank of atomic polarization and is 2 for alignment, $\gamma_1\approx-7\,{\rm{Hz/nT}}$ is the gyromagnetic ratio of the $F=1$ ground state, ${\partial} \phi/{\partial} B|_{\nu_{\rm{m}}=\kappa\gamma_1 B}$ is the  resonant slope of $\phi$ with respect to the magnetic field $B$, which could be read from the $\rm{Y}$ component of the magnetic resonance, and $\delta \phi$ is the sensitivity to the optical rotation angle of the probe beam, measured in ${\rm{rad}/\sqrt{\rm{Hz}}}$.

In the present magnetic resonance scanning, we scan the modulation frequency of the pump beam $\nu_{\rm{m}}$ rather than the magnetic field $B$.
As a result, ${\partial} \phi/{\partial} \nu_{\rm{m}}|_{\nu_{\rm{m}}=\kappa\gamma_1 B}$ can be directly measured from the magnetic resonance data.
Considering the relation $\phi(B,\nu_{\rm{m}})=\phi(B+\Delta B,\nu_{\rm{m}}+\kappa\gamma_1\Delta B)$ when $\Delta B\ll B$, which indicates
\begin{equation}
\label{Eq_MagneticResonanceSlope_1}
\begin{split}
\frac{\partial \phi}{\partial B}
&=\lim_{\Delta B\to 0} \frac{\phi(B+\Delta B,\nu_{\rm{m}}) - \phi(B,\nu_{\rm{m}})}{\Delta B}  \\
&=\lim_{\Delta B\to 0} \kappa\gamma_1\frac{\phi(B,\nu_{\rm{m}}-\kappa\gamma_1\Delta B) - \phi(B,\nu_{\rm{m}})}{\kappa\gamma_1\Delta B}  \\
&=-\kappa\gamma_1\frac{\partial \phi}{\partial \nu_{\rm{m}}},
\end{split}
\end{equation}
we have
\begin{equation}
\label{Eq_MagneticResonanceSlope_2}
\begin{split}
\frac{\partial \phi}{\partial B}\Bigl|_{\nu_{\rm{m}}=\kappa\gamma_1 B}
&=
-\kappa\gamma_1 \frac{\partial \phi}{\partial \nu_{\rm{m}}}\Bigl|_{\nu_{\rm{m}}=\kappa\gamma_1 B} \\
 & = -\kappa\gamma_1 \frac{A}{\delta\nu}  = 2\pi\tau_2 A|\gamma_1|,
 \end{split}
\end{equation}
where $A$ is the optical rotation amplitude on resonance, $\delta\nu$ is HWHM linewidth of the resonance, and $\tau_2=\kappa/(2\pi\delta\nu)$ is the lifetime of atomic polarization.
As a result, the sensitivity is given by
\begin{equation}
\label{Eq_Sensitivity_Estimation_2}
\delta B
\approx
\frac{\delta \phi}{2\pi\tau_2 A|\gamma_1|}.
\end{equation}

%

The photon-shot-noise-limited angular sensitivity $\delta \phi$ is estimated to be $1/\sqrt{P}$~\cite{Budker2000Sensitive}, where $P$ is the photon flux entering the polarimeter. In the present experiment, the power of the 780 nm probe beam entering the vapor cell is about 300\,$\mu{\rm{W}}$, and about one third of it is gathered with the polarimeter, so $\delta \phi$ is about $5.1\times10^{-8}\,{\rm{rad}}/\sqrt{\rm{Hz}}$.

\begin{figure}
\centering
\includegraphics[width=1\columnwidth]{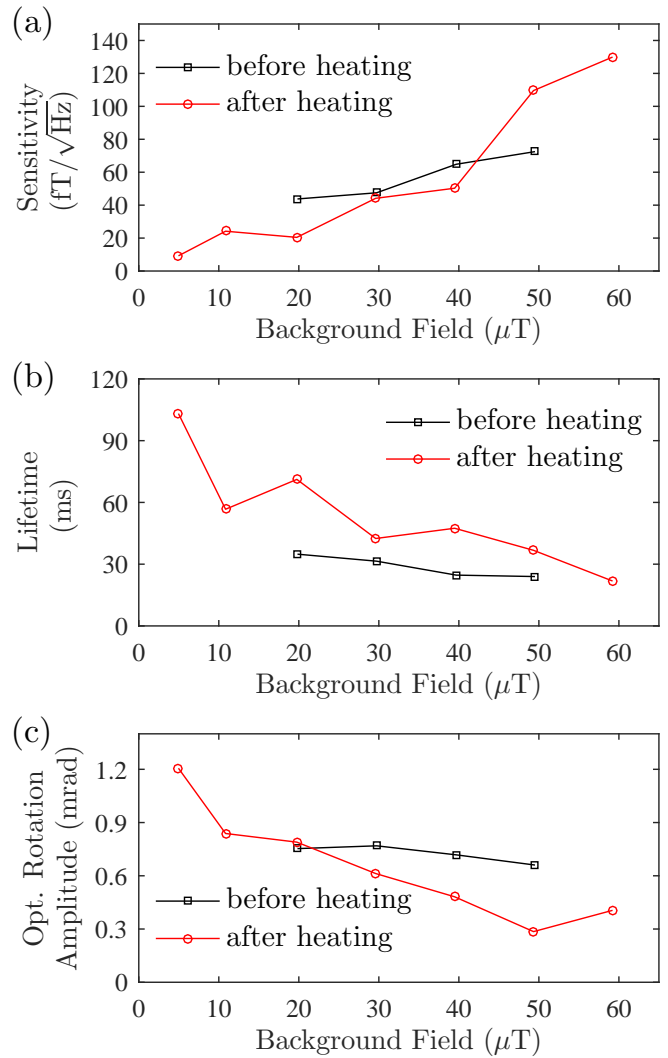}
    \caption{
    \textbf{(a)} Photon-shot-noise-limited sensitivity, \textbf{(b)} Lifetime and \textbf{(c)} optical rotation amplitude of alignment-based magnetometry as a function of background magnetic field.
    The background magnetic fields are kept along the $z$ direction. The black squares (red circles) represent data taken before (after) vapor cell heating.
 }\label{Fig_S1}
\end{figure}

The estimated photon-shot-noise-limited sensitivities of the alignment magnetometry as a function of magnetic field strengths are shown in Fig.~\ref{Fig_S1}(a), with the field kept along the $z$ direction.
We present two sets of results, one of which is measured before the vapor cell heating, and the other is measured after the heating, during which the vapor cell body was heated to about 60$^\circ$C  for several hours and the cell stem was kept as the coldest place of the cell.
The sensitivities are in the tens of ${\rm{fT}}/\sqrt{{\rm{Hz}}}$ range, with the best sensitivity reaching about 9\,${\rm{fT}}/\sqrt{{\rm{Hz}}}$. Generally, the sensitivities degrade with the increased background field. This is possibly due to the increased magnetic field gradient, which in turn decreases the lifetime and amplitude of atomic polarization and thus the optical rotation amplitude, as shown in Fig.~\ref{Fig_S1}(b) and (c).
The overall performance of the magnetometer is improved after heating, which is due to the increased spin lifetime~\cite{Li2017Characterization}.
These results suggest that such an alignment resonance is promising for highly sensitive magnetometry in the geomagnetic field range.

\bibliography{main.bib}

\end{document}